\documentclass[aps,prb,twocolumn]{revtex4}
\usepackage{amssymb,amsbsy,amsmath,graphicx}
\usepackage[export]{adjustbox}

\def\unue#1{{\it #1:}}

\begin{document}

\title{Analytical theory for proton correlations in common water ice $I_h$}

\author{S. V. Isakov$^1$, R. Moessner$^2$, S. L. Sondhi$^{2,3}$, D. A. Tennant$^4$}

\affiliation{$^1$ Google, Brandschenkestrasse 110, 8002 Z\"urich, Switzerland\\ 
$^2$ Max-Planck-Institut f\"ur Physik komplexer Systeme, Dresden\\
$^3$ Dept.\ of Physics, Princeton University, Princeton\\ 
$^4$ Neutron Sciences Directorate, ORNL, Oak Ridge TN 37831}

\date{\today}

\begin{abstract}
We provide a fully analytical
microscopic theory for the proton correlations in water ice $I_h$. We
compute the full diffuse elastic neutron scattering structure factor,
which we find to be in excellent quantitative agreement with Monte
Carlo simulations. It is also in remarkable qualitative agreement with
experiment, in the absence of {\it any} fitting parameters. 
Our
theory thus provides a tractable analytical starting point to account
for more delicate features of the proton correlations in water ice.
In addition, it directly determines an effective field theory of water
ice as a topological phase.
\end{abstract}

\maketitle

%------------------------------------------------------------------------------

%\section{Introduction}
The different phases of matter are commonly illustrated through the
example of ice, water and steam. However, common water ice has for a long
time been known to be a most
untypical solid,\cite{petrwhit} 
exhibiting Pauling's celebrated 'zero-point entropy'\cite{paulingentropy}
as the protons in fact remaining intricately partially disordered, with only the oxygens
being associated with a regular arrangement on a lattice. 

\begin{figure}[ht]
%\includegraphics[width=0.95\columnwidth]{ice_dh_1}
%\includegraphics[width=0.5\columnwidth]{ice_dh_oonly_grab}\\
%\includegraphics[width=0.35\columnwidth]{icearrow}
%{\includegraphics[width=0.3\columnwidth]{iceunit_grab}}
%\includegraphics[width=0.32\columnwidth]{fig1a}
%\includegraphics[width=0.32\columnwidth]{fig1b}
%{\includegraphics[width=0.25\columnwidth]{fig1c}}
\adjincludegraphics[width=0.45\columnwidth,trim={{.25\width} {0.14\height} {.23\width} {0.1\height} },clip]{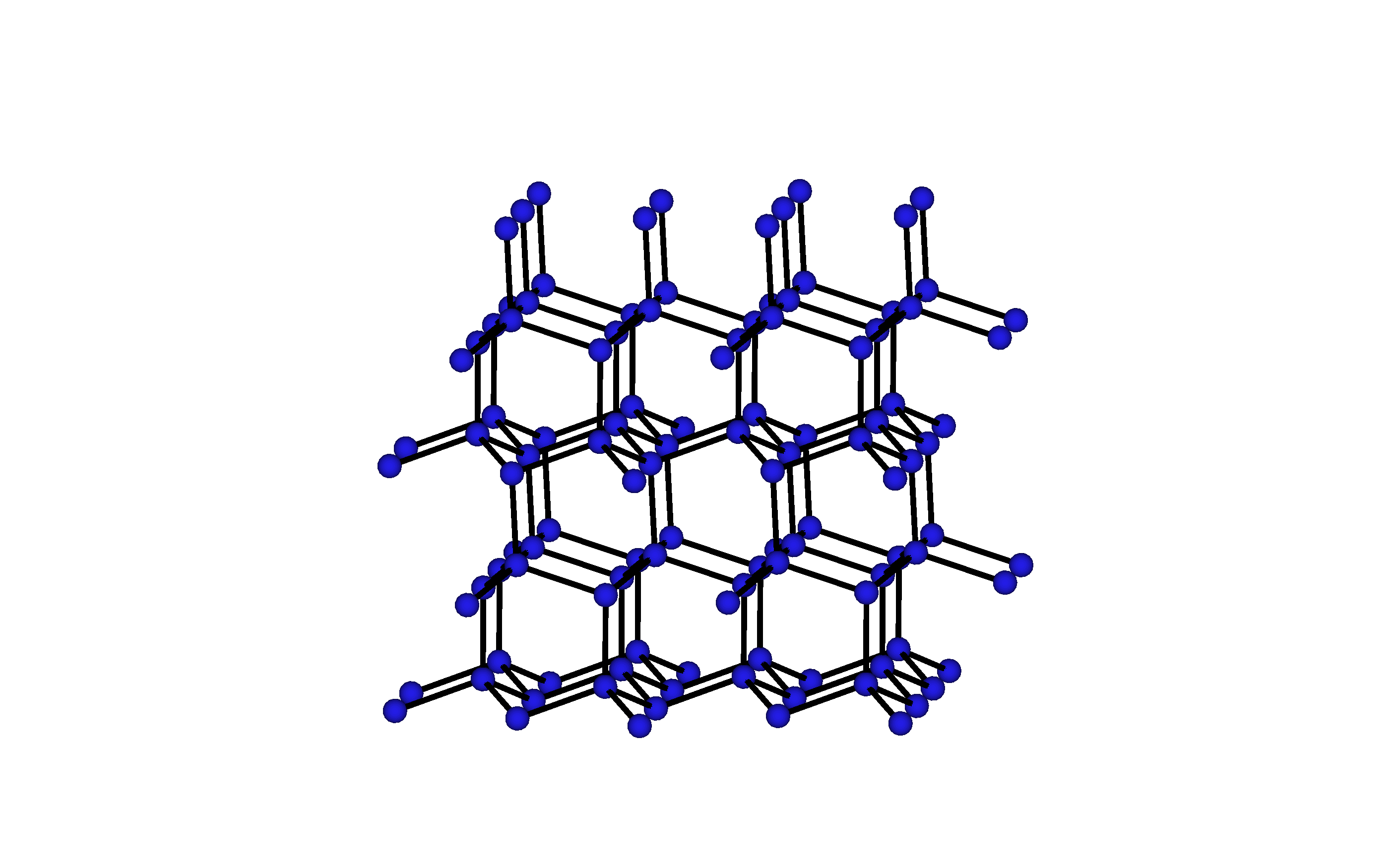}
\includegraphics[width=0.3\columnwidth]{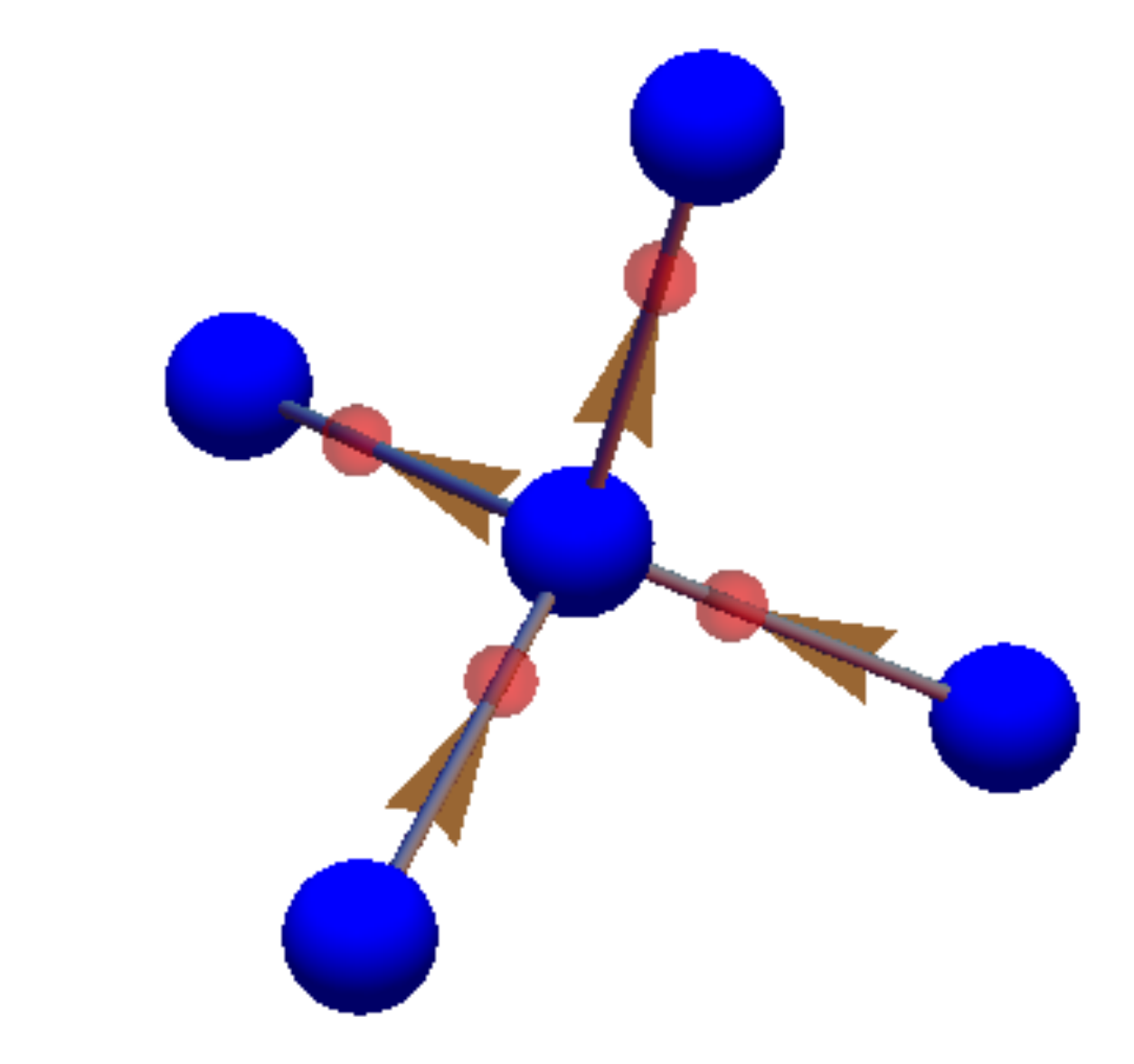}
%\hfill
{\includegraphics[width=0.22\columnwidth]{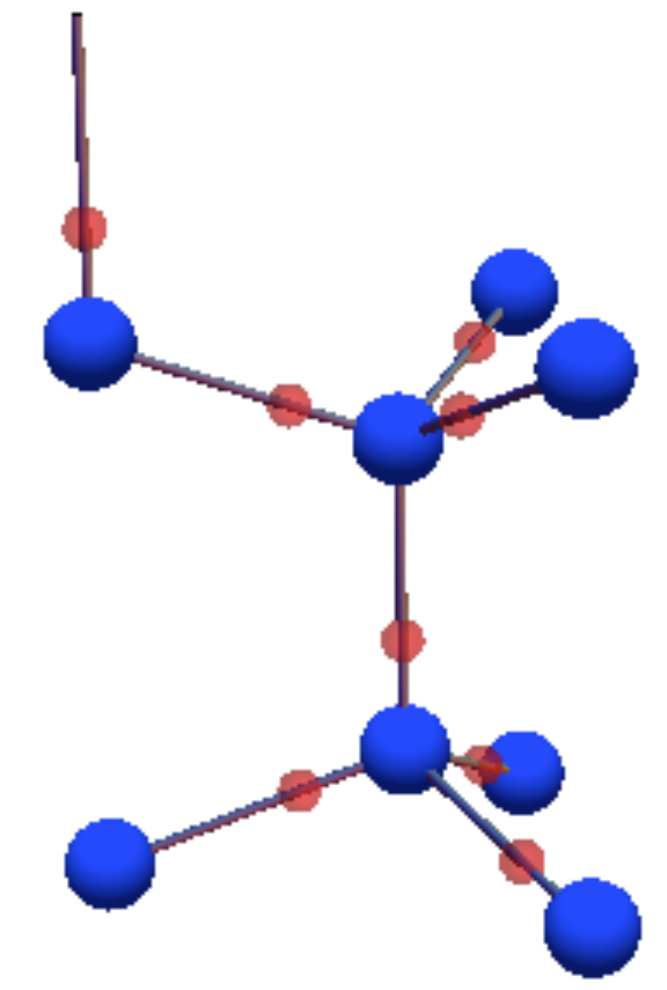}}

\caption{Protons in an ice crystal. The locations of the  oxygen ions
approximate a bipartite 
wurtzite structure (with identical atoms on the two sublattices,
also known as 'hexagonal diamond').\cite{vesta} 
Like in the case of cubic ice, where they form a (cubic) diamond lattice, they 
are four-fold coordinated with the midpoints of the bonds forming tetrahedra. 
However, the tetrahedra are arranged differently:
the relation between these two lattice types is the same as  between 
hexagonal and cubic close packed crystal systems
The oxygens are bonded by the red hydrogen/deuterium ions. The latter are not ordered but 
are subject to the ice rule that each oxygen has two sitting closeby and two further away,
so that the $\{$D/H$\}$$_2$O molecules  `retain their identity' even in the solid. This
implies an emergent conservation law for the flux denoted by the brown arrows on the midpoints
of the bonds, defined to point towards the proton on the bond (middle): each oxygen ion sees the same 
amount of flux arriving as leaving. Right: sketch of a unit cell with an example of an allowed 
configuration of the eight 
protons it contains.}
\label{fig:icecrys}
\end{figure}

Nowadays, we can address detailed questions about
the microscopic nature of water ice. In its most common form, ice $I_h$, the fourfold coordinated 
oxygen atoms form a hydrogen-bonded 
wurtzite structure.\cite{petrwhit}  
One of the  
Bernal-Fowler ice rules state that each oxygen atom has two protons sitting close to it, and   the
other two are further  away (Fig.~\ref{fig:icecrys}). Being subject to these ice rules, proton locations are thus not entirely random, 
and the nature of correlations between them has been a topic of study for a
long time, in theory, simulation and experiment\cite{petrwhit,fradice,expice,mcice,qu_ice,ice_gren}.
In this context, the fact that an ice rule can be cast as a conservation law 
was identified as leading to a characteristic pinch-point feature in the 
structure factor of ice.\cite{villainice,YoungAxe,henleycoulomb} 

However, beyond this long-wavelength insight, 
little analytical progress has occurred, thus providing only limited 
analytical backup  for 
the existing extensive numerical modelling, which does exhibit 
satisfactory agreement with experiment.

In recent years, much progress in this direction has occurred in the context 
of the family of magnetic compounds
known as the spin ices\cite{huse,hermele,pyrdip,henley,qu_ice}, for reviews
see Refs.~\onlinecite{henleycoulomb,cmsARCM} These have benefitted much from 
transfer from water ice -- including receiving their name. In this work, we
aim to reverse the direction of this transfer. This partially builds on insights gained 
in considering spin ice
as a model system for topological phases in condensed matter
physics -- with water ice probably is the most ubiquitous material exhibiting topologically
nontrivial behaviour.

In the following, we first present our analytical theory of the proton correlations in 
water ice, which we contrast to simulations and experiment. We then briefly put this in the context 
 of topological phenomena in condensed matter physics, in particular by 
deriving an accurate value of 
the coupling constant of the long-wavelength action for water ice $I_h$.

%------------------------------------------------------------------------------
\unue{Structure factor} We consider the diffuse scattering away from the Bragg peaks, as the peaks are
encode the `average' underlying arrangement of the ions rather than the displacement of the protons from it, 
which is what we are interested in. 
Whereas X-ray scattering is sensitive
to the charge of the ions and their electrons, and thus is not ideally suited for detecting protons, 
neutrons are a suitable probe, especially for deuterated ice, D$_2$O. 

Our theory neglects
thermal fluctuations and any static disorder that may be present in a given sample and focuses 
on implementing the abovementioned ice rules
%, in particular that each oxygen have two protons sitting close to it, and two further away 
%(Fig.~\ref{fig:icecrys}). 
The basic degree of freedom is thus the location $\mathbf{r}_{i\alpha}$ 
of the proton on bond $\alpha$ of unit cell $i$. We allow two possible proton positions
for each bond -- closer to either one or the other oxygen being bonded. This is
parametrised by an Ising variable $S=\pm1$:
\begin{equation*}
 \mathbf{r}_{i\alpha}=\mathbf{r}^0_{i\alpha}+S_{i\alpha} a
  \hat{\mathbf{e}}_{\alpha},
\end{equation*}
where $\mathbf{r}^0_{i\alpha}$ are the midpoints of the oxygen-oxygen bonds
forming a network of corner-sharing tetrahedra, $S_{i\alpha}$ are the
Ising ``spins'' living at the midpoints of the oxygen-oxygen bonds, $a$ is
the absolute value of the proton displacement relative to the midpoint, and
$\hat{\mathbf{e}}_{\alpha}$ are unit vectors along the oxygen-oxygen bonds.
The index $\alpha$ runs from 1 to 8 as the unit cell of the wurtzite structure
contains 8 sites.

The ice rules are enforced by devising
a Hamiltonian which penalises configurations that do not obey them, and studying the 
resulting ground state 
correlations. How to obtain this has been established in the context of
spin ice \cite{pyrdip}. The precise form of the Hamiltonian, 
with details of the analysis sketched below, is relegated to the appendix. 

The quantity we compute is the structure factor, which is proportional
to the neutron scattering cross section: 
$$
 {\cal S}(\mathbf{q}) = \langle | \sum_{i\alpha}
  b_{i\alpha} e^{i\mathbf{q}\mathbf{r}_{i\alpha}} |^2 \rangle,
$$
where $\mathbf{q}$ is the
wave vector change of the neutron, $b_{i\alpha}\equiv b=\text{const}$ is
the scattering length, and the  brackets denote averaging over all
proton configurations obeying the ice rules. 
\begin{figure}[ht]
\includegraphics[width=0.99\columnwidth]{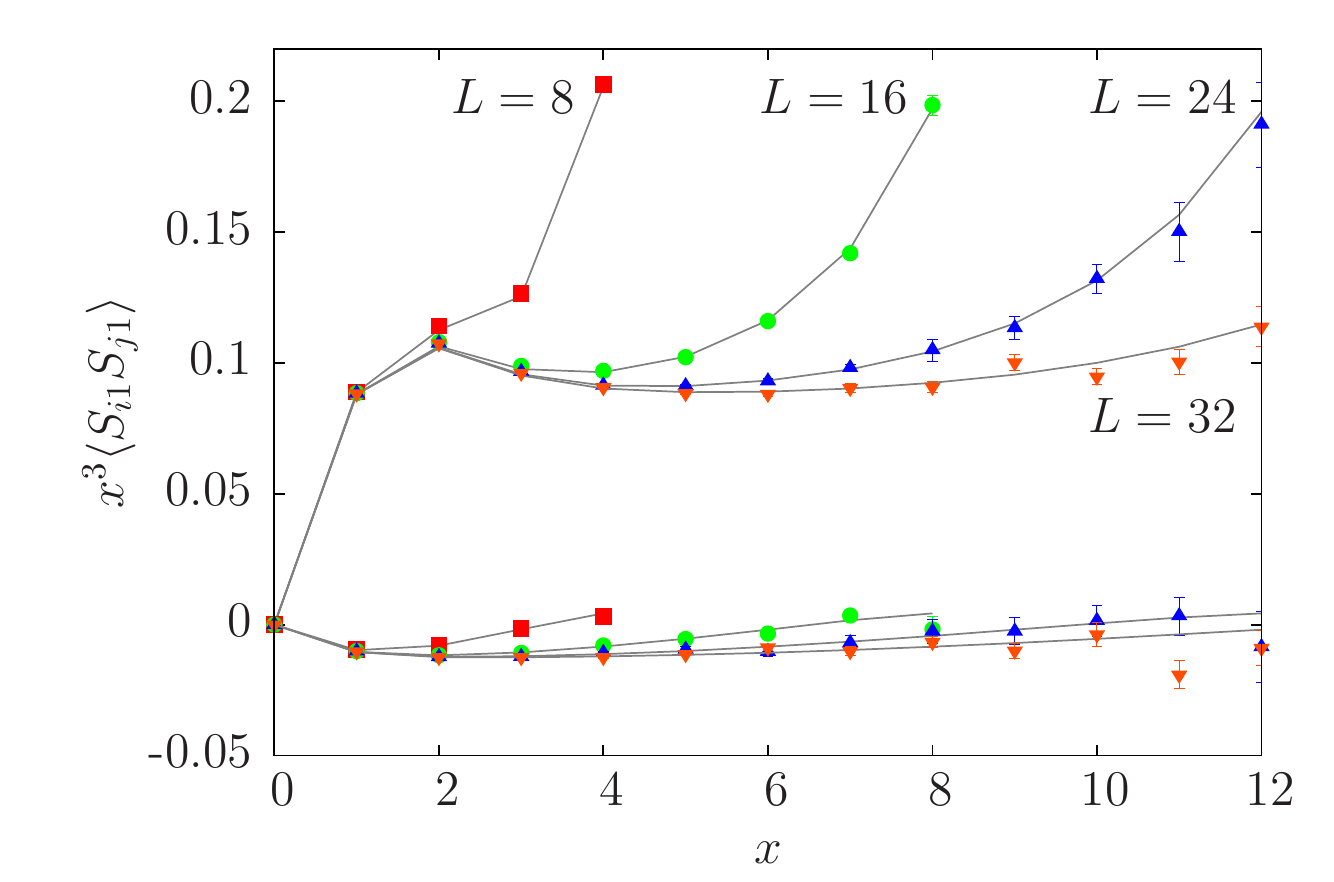}
\caption{Theory versus Monte Carlo simulations. 
Correlation function $\langle S_{i1}S_{j1}\rangle$
from Monte Carlo simulations (symbols) compared to our analytical theory
(lines) at zero temperature for two inequivalent directions, $[100]$ and $[001]$,
and different system sizes $L$.
The correlation functions are multiplied by the cube of the distance $x$ for better
visibility.}
\label{fig:cf:comp}
\end{figure}
To express the structure factor in terms of the Ising
spin correlation function $\langle S_{i\alpha}S_{j\beta} \rangle$, we note
\begin{equation*}
 {\cal S}(\mathbf{q}) = b^2 \sum_{i\alpha,j\beta} \langle
  e^{ia\mathbf{q}(S_{i\alpha}\hat{\mathbf{e}}_{\alpha}
    -S_{j\beta}\hat{\mathbf{e}}_{\beta})}
  \rangle e^{i\mathbf{q}(\mathbf{r}^0_{i\alpha}-\mathbf{r}^0_{j\beta})}.
\end{equation*}
The spin correlator is absent in one of the following terms
\begin{eqnarray*}
 \langle
  e^{ia\mathbf{q}(S_{i\alpha}\hat{\mathbf{e}}_{\alpha}
    -S_{j\beta}\hat{\mathbf{e}}_{\beta})}
 \rangle &=& 2 \langle S_{i\alpha}S_{j\beta} \rangle
   \sin(a\mathbf{q}\hat{\mathbf{e}}_{\alpha})
   \sin(a\mathbf{q}\hat{\mathbf{e}}_{\beta}) \\
  &+& 2 \cos(a\mathbf{q}\hat{\mathbf{e}}_{\alpha})
   \cos(a\mathbf{q}\hat{\mathbf{e}}_{\beta})
\end{eqnarray*}
whence we finally get the diffuse scattering as
\begin{equation*}
 {\cal S}(\mathbf{q}) \propto \sum_{i\alpha,j\beta}
  \langle S_{i\alpha}S_{j\beta} \rangle
  \sin(a\mathbf{q}\hat{\mathbf{e}}_{\alpha})
  \sin(a\mathbf{q}\hat{\mathbf{e}}_{\beta})
  e^{i\mathbf{q}(\mathbf{r}^0_{i\alpha}-\mathbf{r}^0_{j\beta})}.
\end{equation*}
with the computation of the correlation function $\langle S_{i\alpha}S_{j\beta}\rangle$
carried out in the large-$N$ framework (described in the appendix). This is based
on treating relaxing the fixed spin length constraint to one which is only obeyed
`on average', but which has the advantage of being exactly soluble. 
For models closely related to the present ones, this has been 
shown to be an accurate approximation.\cite{pyrdip}

Indeed, to demonstrate the accuracy of our evalution of the correlation function
$\langle S_{i\alpha}S_{j\beta} \rangle$, we compare Monte Carlo simulations
to the analytical expression in Fig.~\ref{fig:cf:comp}. We use the notation for a unit cell
containing four oxygen atoms and eight protons. One finds excellent quantitative
 agreement  in different directions, for different distances,
even capturing finite-size effects faithfully. The data for finite-size systems with $N_t$ spins
can be obtained by explicitly solving for an $N_t\times N_t$-dimensional 
interaction matrix in real space, or equivalently  carrying out a discrete sum over points in
reciprocal space.  
%------------------------------------------------------------------------------

\begin{figure}[ht]
{\includegraphics[width=0.99\columnwidth]{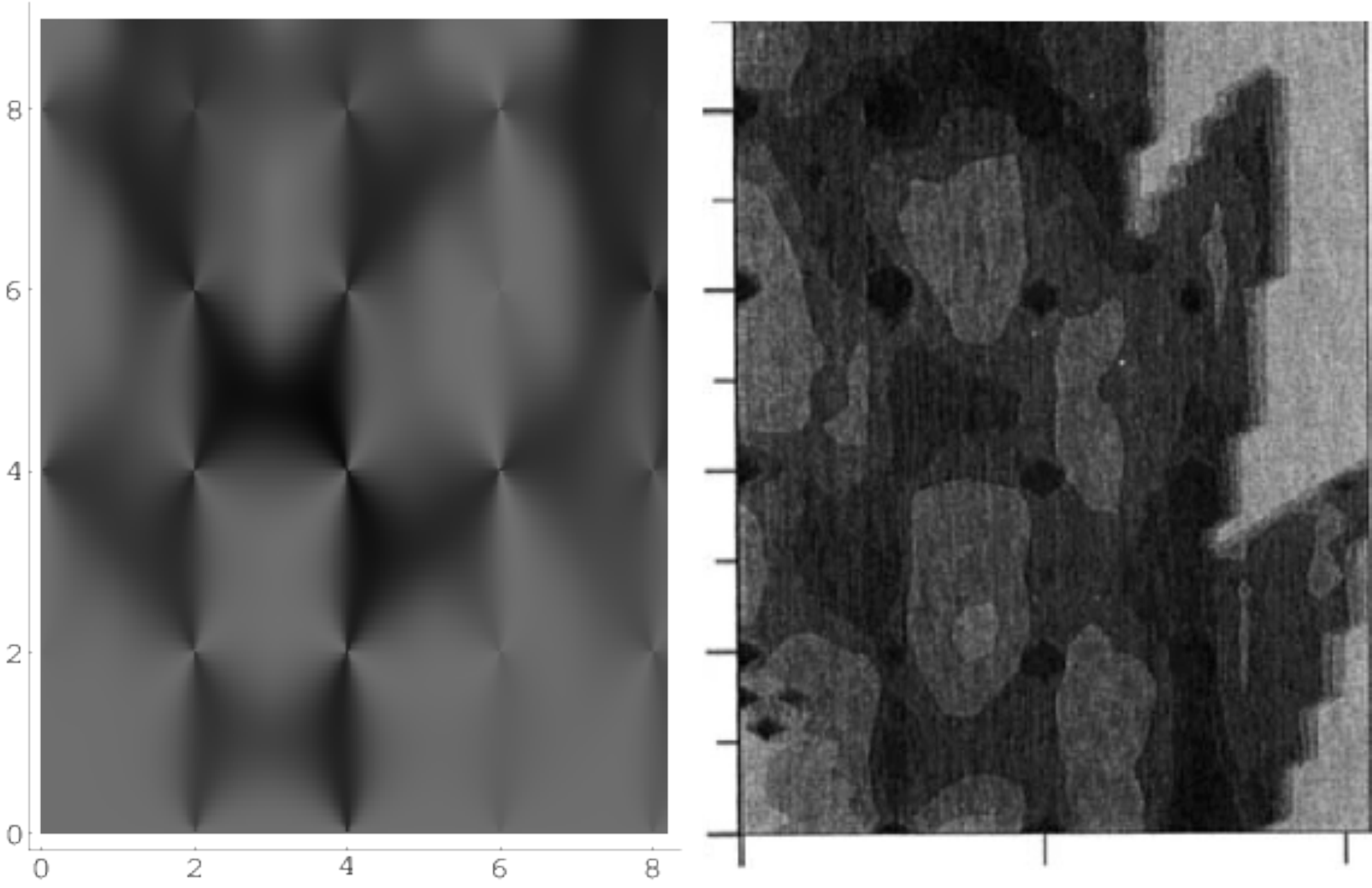}}
{\includegraphics[width=0.99\columnwidth]{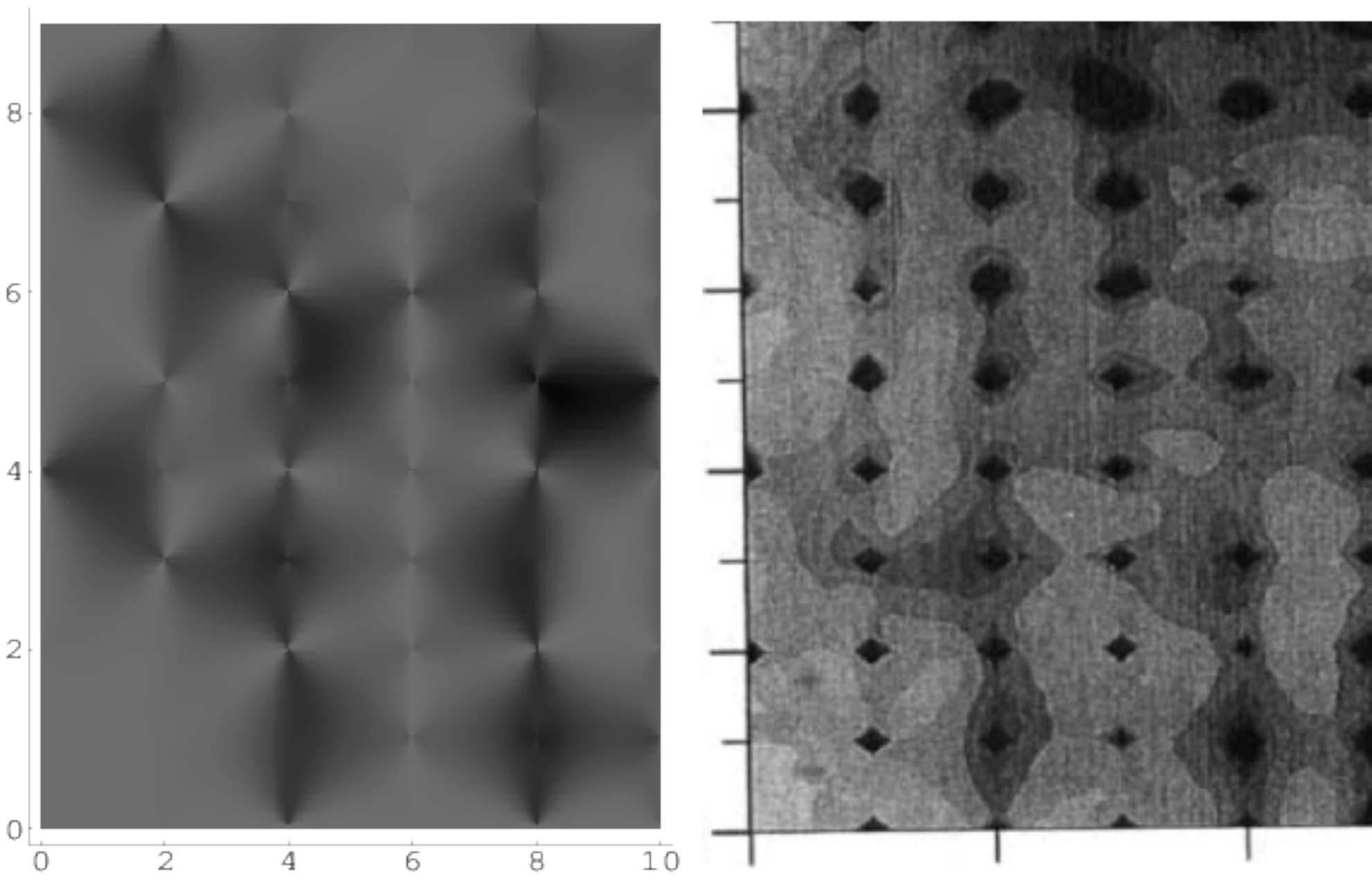}}
\centerline{\includegraphics[width=0.99\columnwidth]{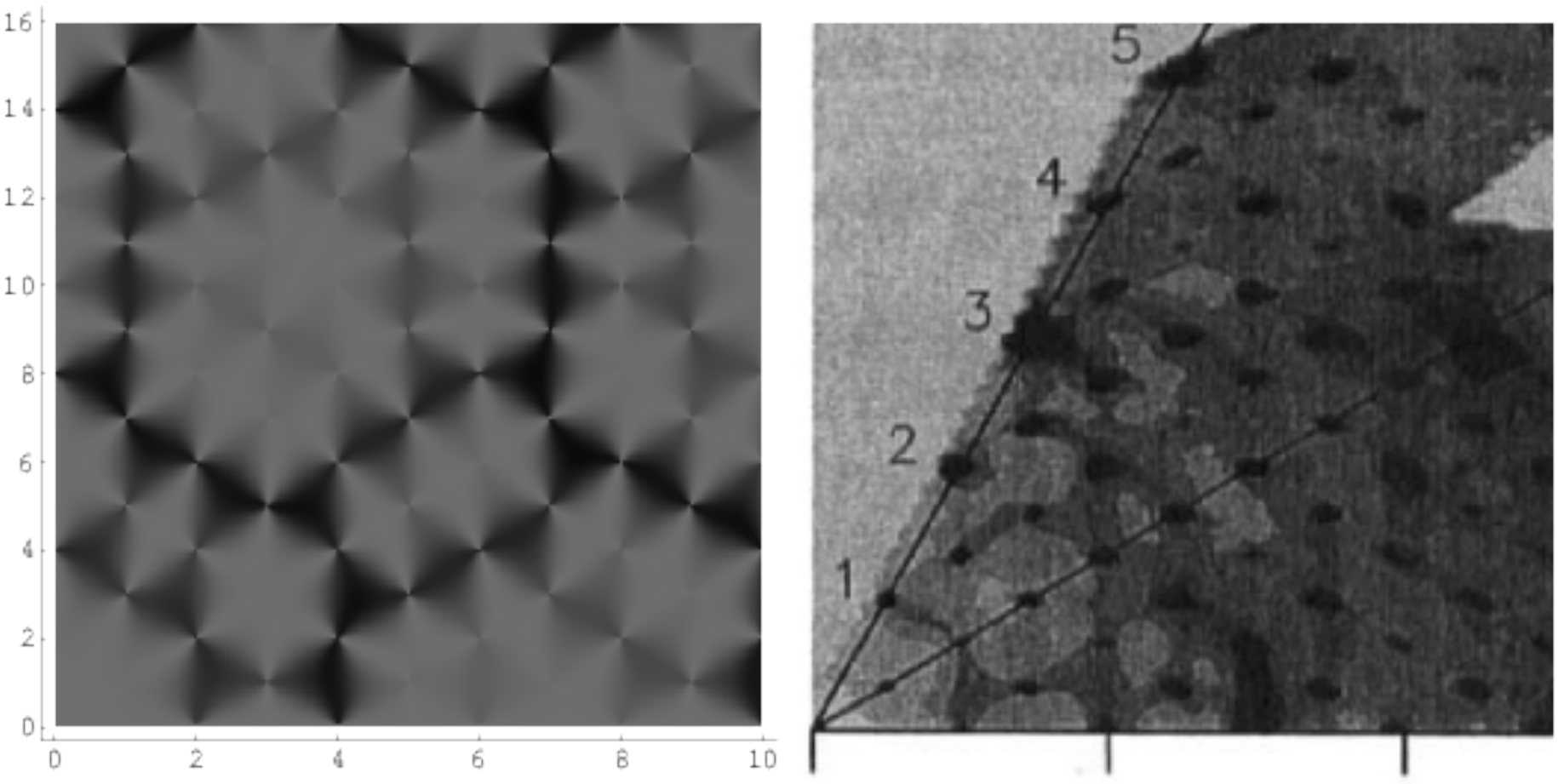}}
\caption{Diffuse neutron scattering in water ice. 
Shown is the structure factor 
in the $[h0l]$ (top), $[0kl]$ (middle) and $[hk0]$ (bottom) planes, with dark  (light)
regions indicating high (low) scattering intensity.\cite{fn-convention}
Left: theoretical results for zero temperature (bottom: Eq.~\ref{eq:sf:hk0}).
Right: experimental result from D$_2$O ice \cite{expice} at 20 K (bottom: 10K). Note the 
pinchpoint features at the centres of higher Brillouin zones, where areas 
of high and low scattering intensity meet; these are partially obscured by Bragg peaks in the experimental
data.}
\label{fig:h0l}
\end{figure}

\unue{Comparison to experiment} The analytically obtained neutron scattering structure factor
is compared to neutron results\cite{expice} in different planes in reciprocal space in 
Figs.~\ref{fig:h0l}.
The proton displacement was taken to be $a\approx 0.1436R_\text{OO}$\cite{petrwhit},
where $R_\text{OO}$ is the distance between the oxygen atoms. 
%The different axes labellings are due to differing choices of coordinate system.

Analytical
results compare reasonably well to the experiment, especially in the first few
Brillouin zones. In particular, location and orientation of the pinchpoints are 
given correctly, alongside the general structure of regions with high and low 
neutron scattering intensity. 

One important feature of our theory is thus that there exists an analytical expression for the 
correlations which can be used as a starting point to understand deviations away from the
ideal ice model. By providing an explicit analytical form, this 
obviates a complex first modelling step  (e.g.\ via numerical fitting
procedures to Monte Carlo simulations) and therefore allows for a focus on the physics
beyond the ice rules. 

Such deviations can take many forms. One is a simple improvement of our
modelling of the scattering form factor of the proton/deuterium ion, which we have treated
as a pointlike scatterer so far, but whose finite extent and finite-temperature thermal
motion will inevitably lead to a change in the 
structure factor in higher Brillouin zones irregardless of any cooperative physics. 
Similarly, 
we have completely omitted the Bragg peaks due to the oxygen ions, which feature 
prominently in the experimental plots. 

More interesting are defects in the lattice structure or 
in the bonding -- violations of the ice rule -- which 
may be static or dynamical. Defects in ice have attracted considerable 
attention, and much is known about them. In our treatment, there is one natural parameter 
which can be added to account for the presence of gauge-charged defects (see below), 
namely an effective correlation length beyond which their presence
removes the correlations responsible for the pinch-points, with the resulting width of the 
pinchpoints reflecting the inverse of this length.

\unue{Long-wavelength theory}
The great advantage of our method lies in the fact that it works 
for all distances as well as a wide variety of lattices. The peculiar 
nature of the proton correlations become particularly evident from the analytics
upon closer inspection, with the 
pinch-point features in the structure factor visible even in the absence of 
an explicit identification of an emergent gauge field. However, historically,
it was noticed already early on\cite{villainice,YoungAxe} that the pinchpoints in the
neutron scattering were due to this conservation law. Basically, the reason this happens
is because a field obeying only the constraint  $\nabla\cdot{\cal B}=0$, but which is otherwise
free,  has no longitudinal 
degrees of freedom, and a pinchpoint is precisely the result  of projecting out the
longitudinal component of the (otherwise random) proton displacements, in a way
which is entirely analogous to how the pinchpoints arise in spin ice \cite{huse,henley}.
[The field ${\cal B}$ is obtained straightforwardly by drawing an arrow 
from the midpoint of the bond towards the centre of charge
of the proton. Upon identifying these arrows with a unit of an imaginary (lattice) flux, ${\cal B}$,the
(lattice) divergence of this flux vanishes, $\nabla\cdot{\cal B}=0$ -- the flux is conserved 
(Fig.~\ref{fig:icecrys}).]

Via an analogy to magnetostatics, one obtains 
an effective partition function, where  the permeability $\mu_0$ of the electromagnetic vacuum is 
replaced by an emergent permeability ('stiffness'), $K$:
$$
{\cal S}=\frac{K}{2} \int d^3 r {\cal B}^2\ . 
$$
This is called topological because the above functional 
goes along with the deconfined `Coulomb' phase of a classical U(1) gauge theory,
which is distinct from a simple disordered phase, while not exhibiting any 
conventional (crystalline) order for the protons.\cite{cmsARCM}

As the wurtzite lattice is not cubic (unlike the case of
spin ice, which corresponds to ice $I_c$), it is not {\it a priori} obvious that 
a single stiffness constant is enough to describe the action of the gauge field, 
but it turns out that just one
stiffness constant is enough, at least approximately.  Indeed, the actual value of this 
stiffness is  a priori a free parameter, as it cannot be derived by symmetry considerations alone,
instead depending on non-universal details of the model exhibiting an emergent
gauge field. Our lattice-based microscopic theory does incorporate such information,
so that it can be used to extract an approximate but accurate estimate from its
long-wavelength expansion near the pinchpoints. For instance, one can expand
Eq.~\ref{eq:sf:hk0} in the Appendix near $q_x=q_y=0$ and read off, by direct
comparison to the predictions from the gauge theory, which have the same functional
form:
\begin{equation}
K R_\text{OO}^3 \approx \sqrt{3}/8\ .
\label{eq:kstiff}
\end{equation}
This piece of information, together with the general long-wavelength form of the theory, 
is enough to yield the full asymptotics of the 
correlation function in real space, which takes  the dipolar form,
$$
 \langle S_{i\alpha}S_{j\beta} \rangle = \frac{1}{4\pi K} \left(
  \frac{\hat{\mathbf{e}}_{\alpha}\hat{\mathbf{e}}_{\beta}}{r_{ij}^3}
  	- \frac{3(\hat{\mathbf{e}}_{\alpha}\mathbf{r}_{ij})
	   (\hat{\mathbf{e}}_{\beta}\mathbf{r}_{ij})}{r_{ij}^5}
 \right).
$$
 
Note that these results imply that the long-range part of the interaction between
the protons -- which is due to the interactions between the dipole moments of
the bonds arising from the asymmetric location of the protons -- has an additional
component to it which is {\it not} of an electrostatic origin, but rather due
to the contribution Eq.~\ref{eq:kstiff}. This would be present even if the
protons were electrically neutral, so that neutral particles obeying the ice rules
would exhibit the same form of the correlations!

This dichotomy is due to the protons being doubly charged--they have an (intrinsic) 
electric on top of an (emergent) gauge charge. The latter associated with 
the emergent conservation law for ${\cal B}$. The former is not the naive `electronic' 
charge $|e|$, but rather related to the divergence of the electric dipole moment
at the location of a defect.\cite{NagleUnit,MoeSonIrrat} Charged defects in ice are therefore
special in that they are quasiparticles which not only have 
an {\it irrational electric} charge, but also 
an emergent {\it entropic} Coulomb charge. 

\unue{Final remarks}  It is remarkable that a material as well-known as
water ice should be an embodiment of the topological physics of unconventional
types of order that has come into focus relatively recently.
For future work, it would be most desirable to obtain new neutron scattering
data on a par with the recent X-ray work,\cite{ice_gren} in order to allow for a more detailed
and quantitative understanding of the microscopic proton distribution in water ice,
beyond the analysis presented there; here
we have provided a parameter-free analytical theory encoding the Bernal-Fowler ice rules, 
which is applicable at all lengthscales and can be used as a solid basis for including
more delicate effects in a simple framework. It is a tantalising prospect that such analytical 
theories might be more generally useful for modelling diffuse neutron scattering experiments.

\unue{Acknowledgements}
This work was in part supported by the Helmholtz Virtual Institute VI-521
"New states of matter and their excitations" (R.M.). This work was supported
by NSF Grant No. DMR-1311781, the Alexander von Humboldt Foundation and
the German Science Foundation (DFG) via the Gottfried Wilhelm Leibniz Prize
Programme at MPI-PKS (S.L.S.). ORNL is managed by UT-Battelle, LLC, under
contract DE-AC05-00OR22725 for the U.S. Department of Energy (D.A.T.).

\unue{Note added} At the same time as this work, a preprint by Benton et al.\ appeared\cite{BentonIce} 
on the proton correlations in water ice, in particular in the presence of quantum dynamics.

\appendix
\section{Large-$N$ theory for water ice $I_h$}
The (approximate) oxygen positions in water ice can be used to define a 
regular wurtzite structure with a unit cell containing four oxygen ions, to 
which contains 8 protons (spins) are associated. The 
effective interaction matrix enforcing the ice rule therefore has dimension
$8\times8$ and takes the form
$$
{\cal H}=
\frac{1}{T} \left(
\begin{matrix}
  2   & 2b       & 2c       & f         & f^*     & 0         & 0         & 0         \cr 
  2b  & 2        & 2\bar{c} & g         & g^*     & 0         & 0         & 0         \cr
  2c  & 2\bar{c} & 2        & h         & h^*     & 0         & 0         & 0         \cr
  f^* & g^*      & h^*      & 2         & 0       & \bar{f}   & \bar{g}   & \bar{h}   \cr
  f   & g        & h        & 0         & 2       & \bar{f}^* & \bar{g}^* & \bar{h}^* \cr
  0   & 0        & 0        & \bar{f}^* & \bar{f} & 2         & 2b        & 2c        \cr
  0   & 0        & 0        & \bar{g}^* & \bar{g} & 2b        & 2         & 2\bar{c}  \cr
  0   & 0        & 0        & \bar{h}^* & \bar{h} & 2c        & 2\bar{c}  & 2         \cr
\end{matrix}
\right),
$$
where
$b=\cos(q_x/2\sqrt{2})$,
$c=\cos(q_x/4\sqrt{2}+\sqrt{3}q_y/4\sqrt{2})$,
$\bar{c}=\cos(q_x/4\sqrt{2}-\sqrt{3}q_y/4\sqrt{2})$,
$f=\exp[i(q_x/4\sqrt{2}+q_y/4\sqrt{6}+q_z/2\sqrt{3})]$,
$g=\exp[i(-q_x/4\sqrt{2}+q_y/4\sqrt{6}+q_z/2\sqrt{3})]$,
$h=\exp[i(-q_y/2\sqrt{6}+q_z/2\sqrt{3})]$,
$\bar{f}=\exp[i(-q_x/4\sqrt{2}-q_y/4\sqrt{6}+q_z/2\sqrt{3})]$,
$\bar{g}=\exp[i(q_x/4\sqrt{2}-q_y/4\sqrt{6}+q_z/2\sqrt{3})]$,
$\bar{h}=\exp[i(q_y/2\sqrt{6}+q_z/2\sqrt{3})]$, $*$ denotes complex conjugation,
and $T$ is the temperature.

The Fourier transform of the Ising spin correlation function
$G_{\alpha\beta}(x)=\langle S_{i\alpha}S_{j\beta}\rangle$ can be written
in the large-$N$ theory approximation as\cite{garcan,pyrdip}
$$
 G_{\alpha\beta}(\mathbf{q}) = \sum_{\mu=1}^{8}
  \frac{U_{\mathbf{q},\alpha\mu}U_{\mathbf{q},\beta\mu}^{\dagger}}
       {\lambda + e_{\mathbf{q},\mu}},
$$
where $e_{\mathbf{q},\rho}$ are the eigenvalues of the interaction matrix
${\cal H}$, $U_{\mathbf{q},\alpha\mu}$ is a unitary transformation that
diagonalizes the interaction matrix, and $\lambda$ is determined by
the equation
$$
8n=\sum_{\mathbf{q},\mu} \frac{1}{\lambda + e_{\mathbf{q},\mu}},
$$
where $n$ is the number of unit cells. In the limit of  low temperature,
only the four flat bands of the interaction matrix contribute to
the correlation function and $\lambda\rightarrow 1/2$. 

The structure factor can then be straightforwardly obtained analytically, although it turns out that its form is  rather lengthy.
In some high-symmetry planes, it however, becomes quite simple. For instance,  in the $[hk0]$ plane 
(with ${\cal C}_x=\cos \left(\frac{a q_x}{2 \sqrt{2}}\right)$ and ${\cal C}_y= \cos \left(\frac{a
q_y}{2\sqrt{6}}\right)$ and ${\cal C}\leftrightarrow{\cal S}$ for $\cos\leftrightarrow\sin$):

\begin{widetext}

\begin{equation}
\frac{\left[
2{\cal C}_x {\cal S}_y \cos \left(\frac{1}{4}
\sqrt{\frac{3}{2}} q_y\right) \sin \left(\frac{q_x}{4 \sqrt{2}}\right) 
-2{\cal C}_y {\cal S}_x\cos \left(\frac{q_x}{4 \sqrt{2}}\right)   \sin
\left(\frac{1}{4} \sqrt{\frac{3}{2}} q_y\right)
+ %\sin \left(\frac{aq_y}{\sqrt{6}}\right)
2 {\cal C}_y {\cal S}_y
\sin \left(\frac{q_x}{2 \sqrt{2}}\right)
\right]^2}
{3-\cos \left(\frac{q_x}{\sqrt{2}}\right)-2 \cos \left(\frac{q_x}{2
\sqrt{2}}\right) \cos \left(\frac{1}{2} \sqrt{\frac{3}{2}} q_y\right)}.
\iffalse
\frac{\left[
2\cos \left(\frac{a q_x}{2 \sqrt{2}}\right) \cos \left(\frac{1}{4}
\sqrt{\frac{3}{2}} q_y\right) \sin \left(\frac{q_x}{4 \sqrt{2}}\right) \sin
\left(\frac{a q_y}{2 \sqrt{6}}\right)
-2\cos \left(\frac{q_x}{4 \sqrt{2}}\right) \cos \left(\frac{a q_y}{2
\sqrt{6}}\right) \sin \left(\frac{a q_x}{2 \sqrt{2}}\right) \sin
\left(\frac{1}{4} \sqrt{\frac{3}{2}} q_y\right)
+\sin \left(\frac{q_x}{2 \sqrt{2}}\right) \sin \left(\frac{a
q_y}{\sqrt{6}}\right)
\right]^2}
{3-\cos \left(\frac{q_x}{\sqrt{2}}\right)-2 \cos \left(\frac{q_x}{2
\sqrt{2}}\right) \cos \left(\frac{1}{2} \sqrt{\frac{3}{2}} q_y\right)}.
\fi
\label{eq:sf:hk0}
\end{equation}
\end{widetext}

%------------------------------------------------------------------------------


\begin{thebibliography}{99}
\bibitem{fn-convention}
Our work and Ref.~\onlinecite{expice} use different conventions for indexing 
the lattice; the figures, however, show the same parts of reciprocal space.
\end{thebibliography}
\end{document}